\documentclass[aps,prl,twocolumn,groupedaddress,showpacs]{revtex4}
\usepackage{graphics}
\usepackage{amsmath}
\usepackage{epsfig}
\usepackage{epsf}
\newcommand {\sla}[1]{ #1 \!\!\!/}

\begin{document}
\title{Contribution of Two-Boson-Exchange with $\Delta(1232)$ Excitation
to\\Parity-Violating Elastic Electron-Proton Scattering}

\date{\today}

\author{ Keitaro Nagata$^1$, Hai Qing Zhou$^2$,
Chung Wen Kao$^1$, and Shin Nan Yang$^{3,4}$ \\
$^1$Department of Physics, Chung-Yuan Christian University,
Chung-Li 32023, Taiwan\\
$^2$Department of Physics,
Southeast University, Nanjing 211189, China\\
$^3$Department of Physics and $^4$Center for Theoretical Sciences,
National Taiwan University,\\
Taipei 10617, Taiwan\\
}
\begin{abstract}
We study the leading electroweak corrections in the precision
measurement of the strange form factors. Specifically, we  calculate
the two-boson-exchange (TBE), two-photon-exchange (TPE) plus $\gamma
Z$-exchange ($\gamma Z$E), corrections with  $\Delta(1232)$
excitation to the parity-violating asymmetry of the elastic
electron-proton scattering.
The interplay between nucleon and $\Delta$ contributions is found to
depend strongly on the kinematics, as $\delta_\Delta$  begins as
negligible at backward angles but becomes very large and negative
and dominant at forward angles, while $\delta_N$ always stays
positive and decreases monotonically with increasing $\epsilon$. The
total TBE corrections to the extracted values of $G^{s}_{E}+\beta
G^{s}_{M}$ in recent experiments of HAPPEX and G0 are, depending on
kinematics, found to be large and range between $13\%$ to $-75\%$
but small in the case of A4 experiments.

\end{abstract}
 \maketitle

\begin{figure}[t]
\centerline{\epsfxsize 1.4 truein\epsfbox{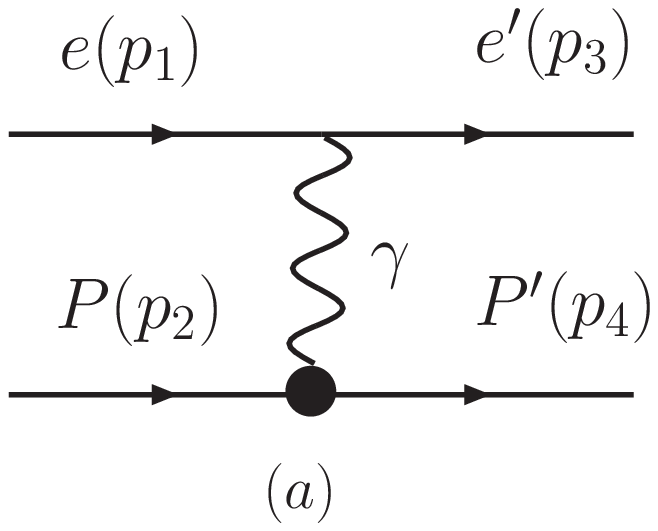}
\epsfxsize 1.4 truein\epsfbox{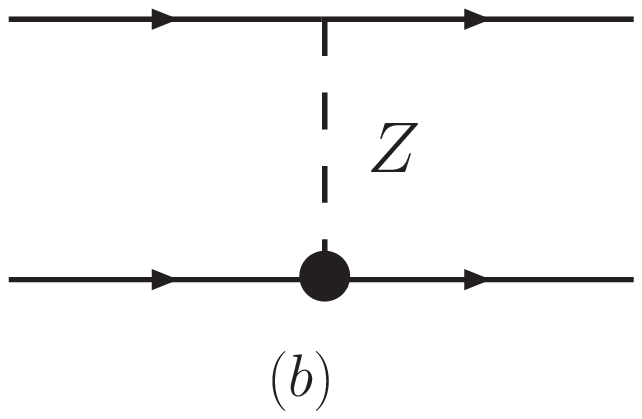}} \centerline{\epsfxsize
1.4 truein\epsfbox{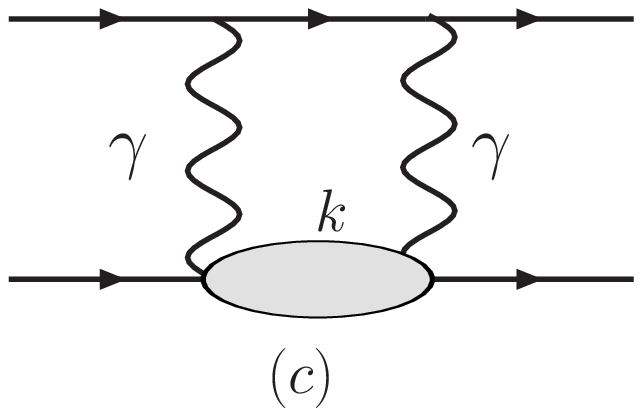} \epsfxsize 1.4
truein\epsfbox{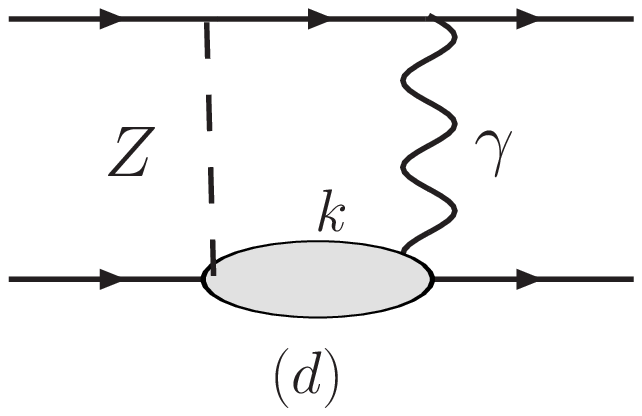}}
\vskip-0.5cm
\caption{(a) one-photon-exchange,
(b) $Z$-boson-exchange, (c) TPE, and (d) $\gamma Z$-exchange
diagrams for elastic {\it ep} scattering. Corresponding cross-box
diagrams are implied.}
\end{figure}

Recently, parity-violating (PV) elastic electron-proton scattering
has been actively pursued \cite{expts} in order to probe the
strangeness content of the nucleon, one of the most intriguing
questions in hadron structure in the last two decades. It is
achieved by measuring the parity-violating asymmetry
$A_{PV}=(\sigma_R-\sigma_L)/(\sigma_R+\sigma_L)$, where
$\sigma_{R(L)}$ is the cross section with a right- (left-) handed
electron, in scattering of longitudinally polarized electrons from
unpolarized protons. At tree level, parity violation in electron
scattering comes from the interference between one-photon-exchange
(OPE) and $Z$-boson-exchange (ZBE) diagrams as shown, respectively,
in Figs. 1(a) and 1(b). The form factors of neutral weak current
extracted from $A_{PV}$ are sensitive to a different linear
combination of the three light quark distributions.  When combined
with proton and neutron electromagnetic form factors and with the
use of charge symmetry, the strange electric and magnetic form
factors, $G^s_E$ and $G^s_M$, can then be determined \cite{Kaplan}.

Leading order radiative corrections to $A_{PV}$, including the box
diagrams  Fig. 1(d) and other diagrams, were extensively studied in
\cite{Marciano83,Musolf90} where the interference between $\gamma
Z$-exchange ($\gamma ZE$) of Fig. 1(d) with  Fig. 1(a), was
evaluated  within the zero momentum transfer approximation, i.e.,
$Q^2=0$. However, recent studies \cite{zhou07,tjon08} indicate that
the $\gamma ZE$ correction has a strong $Q^2$ dependence which could
lead to substantial errors in the extracted strange form factors if
the results obtained with $Q^2=0$ are used.

In addition, the contribution of the interference of the
two-photon-exchange (TPE) process of Fig. 1(c) with diagram of Figs.
1(a) and 1(b) to $A_{PV}$, has been calculated in \cite{afanasev05}
in a parton model using GPDs. It was found that indeed the TPE
correction to $A_{PV}$ can reach several percent in certain
kinematics, becoming comparable in size with existing experimental
measurements of strange-quark effects in the proton neutral weak
current. The results of the partonic calculation of
\cite{afanasev05} at large $Q^2$ have been confirmed by the
 hadronic calculations of \cite{zhou07,tjon08}.

 The calculations of \cite{zhou07,tjon08} which studied the
two-boson-exchange (TBE) corrections to $A_{PV}$, namely, the
contributions of the interference of the TPE process of Fig. 1(c)
with diagram of Figs. 1(a) and 1(b) to $A_{PV}$, and that between
the $\gamma ZE$ of Fig. 1(d) with Fig. 1(a), were carried out in a
simple hadronic model where the intermediate states are restricted
only to elastic intermediate states. Since $\Delta(1232)$ plays a
dominant role in low-energy hadron physics \cite{pascal07}, it is
essential to include the $\Delta$ in the intermediate states and
evaluate the corresponding contribution to the  PV elastic $ep$
scattering. Similar effect, i.e., TPE contribution with an
intermediate $\Delta$ resonance, in the parity-conserving elastic
$ep$ scattering was found \cite{kondra05} to be small as compared
to TPE diagram with nucleon intermediate states. It can be
understood because there is a mismatch in the matrix elements of
$\gamma ee$ and $\gamma N\Delta$ vertices. Namely, the matrix
element of the spatial component of the electron current is
$(v/c)$  smaller than that of the charge component while it is the
other way around in the $\gamma N\Delta$ transition current, i.e.,
the magnetic dipole transition dominates \cite{KY99}. However,
such a mismatch is absent in the diagrams involving $Z$-exchange.
We may then expect $\Delta$ TBE contribution to be non-negligible.

In this paper, we report on calculations of the corrections of the
TBE with $\Delta$ in the intermediate states,
including TPE and $\gamma ZE$, to $A_{PV}$ and their effects on
the extracted values of the nucleon strange form factors.

At hadron level, the photon-induced transition of
$p\rightarrow\Delta^+$ is given by \cite{kondra05}
\begin{eqnarray}
&&\langle N(p')|J^{em}_\mu|\Delta (p)\rangle =\frac{
F_{\Delta}(q^2)}{M_{N}^{2}}\overline{u}(p')
[g_{1}(g^{\alpha}_{\mu}\sla{p}\sla{q}
-p_{\mu}\gamma^{\alpha}\sla{q}\nonumber \\
&&-\gamma_{\mu}\gamma^{\alpha}p\cdot q+\gamma_{\mu}\sla{p}q^{\alpha})
+g_{2}(p_{\mu}q^{\alpha}-p\cdot q g^{\alpha}_{\mu})
+g_{3}/M_{N}\nonumber \\
&&(q^2(p_{\mu}\gamma^{\alpha}-g^{\alpha}_{\mu}\sla{p})
+q_{\mu}(q^{\alpha}\sla{p}-\gamma^{\alpha}p\cdot
q))]\gamma_{5}T_3u_{\alpha}^{\Delta}(p), \label{GamaNDelta}
\end{eqnarray}
where $q=p'-p$ and $T_3$ is the third component of the $N\rightarrow
\Delta$ isospin transition  operator.

The neutral weak current can be decomposed into isovector and
isoscalar parts:
\begin{eqnarray}
J^{Z}_{\mu}&=&\alpha_V V^{3}_{\mu}+\alpha_A A^{3}_{\mu}+\rm{isoscalar}\,\, \rm{terms},\nonumber \\
J^{em}_{\mu}&=&V^{3}_{\mu}+\rm{isoscalar}\,\, \rm{terms},
\end{eqnarray}
where the superscript "3" refers to the 3rd component in isospin
space,  $\alpha_V=(1-2\sin^2\theta_{W})/(2\cos\theta_{W})$, and
$\alpha_A=-1/(2\cos\theta_{W})$. The isoscalar part does not
contribute to N $\rightarrow \Delta$ transition. The vector and
axial-vector components in the $Zp\Delta^{+}$ vertex  can be written
as
\begin{eqnarray}
&&\langle p(p')|J^{Z,V}_\mu|\Delta^{+}(p)\rangle
=\frac{F_{\Delta}(q^2)}{M_{N}^{2}}\overline{u}(p')
[\tilde{g}_{1}(g^{\alpha}_{\mu}\sla{p}\sla{q}
-p_{\mu}\gamma^{\alpha}\sla{q}\nonumber \\
&&-\gamma_{\mu}\gamma^{\alpha}p\cdot q+\gamma_{\mu}\sla{p}q^{\alpha})
+\tilde{g}_{2}(p_{\mu}q^{\alpha}-p\cdot q g^{\alpha}_{\mu})
+\tilde{g}_{3}/M_{N}(q^2\nonumber \\
&&(p_{\mu}\gamma^{\alpha}-g^{\alpha}_{\mu}\sla{p})
+q_{\mu}(q^{\alpha}\sla{p}-\gamma^{\alpha}p\cdot q))]\gamma_{5}u^\Delta_{\alpha}(p),\nonumber \\
&&\langle p(p')|J^{Z,A}_\mu|\Delta^{+}(p)\rangle = \frac{
H_{\Delta}(q^2)}{M_{N}^{2}}\overline{u}(p')
[h_{1}(g^{\alpha}_{\mu}(p\cdot q)-p_{\mu}q^{\alpha})\nonumber \\
&&+h_{2}/M_{N}^{2}(q^{\alpha}q_{\mu}\sla{p}\sla{q}
-(p\cdot q)\gamma^{\alpha}q_{\mu}\sla{q})
+h_{3}((p\cdot q)\gamma^{\alpha}\gamma_{\mu}\nonumber \\
&&-\sla{p}\gamma_{\mu}q^{\alpha})
+h_{4}(g^{\alpha}_{\mu}p^{2}-p_{\mu}\gamma^{\alpha}\sla{p})]u^\Delta_{\alpha}(p),
\label{ZNDelta}
\end{eqnarray}
where $\tilde{g}_{i}'s$ and $g_{i}'s$ are related by
$\tilde{g}_{i}=\sqrt{2/3}\, \alpha_{V} g_{i}.$ $F_\Delta$ and
$H_\Delta$ in Eqs. ({\ref{GamaNDelta}) and (\ref{ZNDelta}}) are
vector and axial-vector form factors, respectively, and we
assume  that, for simplicity,  each of them separately take a
common form for different couplings.

Choosing the Feynman gauge and neglecting the electron mass $m_e$ in
the numerators, the amplitudes of box diagrams of TBE with $\Delta$
excitation as depicted in Figs. 2(a) and  2(b) can be written down
straightforwardly. E.g., we have for Fig. 2(b)

\begin{figure}[h,b,t]
\centerline{\epsfxsize 1.44 truein\epsfbox{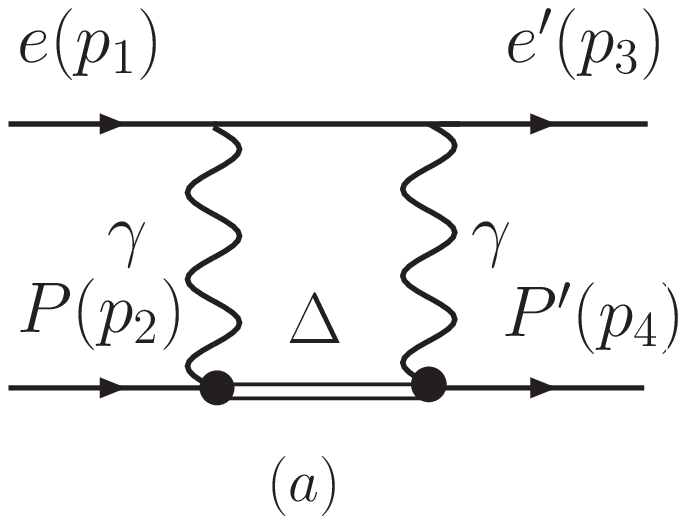} \epsfxsize
1.4 truein\epsfbox{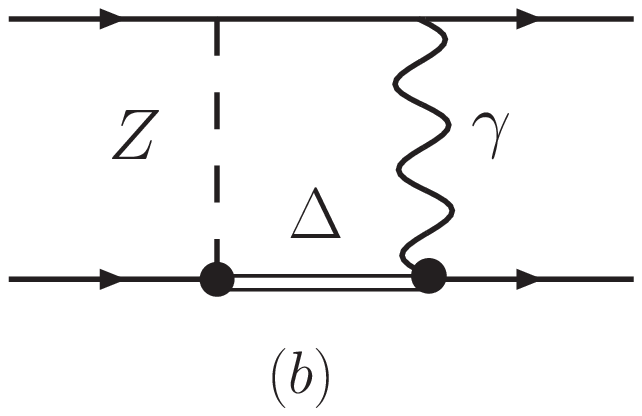}}
\vskip-0.4cm
\caption{(a) TPE and (b) $\gamma ZE$
box diagrams with  $\Delta$ intermediate states. Corresponding
crossed diagrams are implied.} \label{handbag}
\end{figure}

\begin{align}
&M^{(2b)}=-i\int\frac{d^4k}{(2\pi)^4}\overline{u}(p_3)(-ie\gamma_{\mu})\frac{i(\sla{p}_1+\sla{p}_2-\sla{k})}
{(p_1+p_2-k)^2-m_e^2+i\varepsilon} \notag \\
&\times(i\frac{g\gamma_\nu}{4\cos\theta_{W}})[(-1+4\sin^{2}\theta_{W})+\gamma_{5}]u(p_1)
\frac{-i}{(p_4-k)^2+i\varepsilon}\notag \\
&\times\frac{-i}{(k-p_2)^2-M_Z^2+i\varepsilon}
\overline{u}(p_4)\Gamma^{\mu\alpha,\gamma}_{\Delta\rightarrow N}(k,p_{4}-k)\notag \\
&\times\frac{-i(\sla{k}+M_{\Delta})P_{\alpha\beta}^{3/2}(k)}
{k^2-M_{\Delta}^2+i\varepsilon}\Gamma^{\beta\nu,Z}_{N\rightarrow
\Delta}(k,k-p_{2})u(p_2),
\end{align}
where $P_{\alpha\beta}^{3/2}(k)$ \cite{pascal07} is the spin-3/2
projector and $g=e/\sin\theta_{W}$ is the weak coupling constant.
The vertex functions $\Gamma's$ are defined by
$\overline{u}(p')\Gamma^{\mu\alpha,\gamma}_{\Delta\rightarrow
N}(p,q)u^\Delta_{\alpha}(p)=-ie\langle
N(p')|J^\mu_{em}|\Delta(p)\rangle$ and
$\overline{u}^\Delta_{\beta}(p )\Gamma^{\beta\nu,Z}_{N\rightarrow
\Delta}(p,-q)u(p')= -ig\langle \Delta(p)|J^\nu_{Z}|N(p')\rangle$.
Note that the vertices $\gamma N\Delta$ and $ ZN\Delta$ given in
Eqs. ({\ref{GamaNDelta}) and (\ref{ZNDelta}}) both satisfy the
constraints, $ p_{\alpha}\Gamma^{\mu\alpha}_{\Delta\rightarrow
N}(p,q')=p_{\beta}\Gamma^{\beta\nu}_{N\rightarrow \Delta}(p,q')=0$,
for any $q'$ to eliminate the coupling of the unphysical spin-1/2
component of Ratria-Schwinger spinor \cite{Pascal99}. $ZN\Delta$
vertex has been expressed in several other forms
\cite{smith,Nath,Hemmert} which are different from Eq.
({\ref{ZNDelta}). All of them, including Eq. ({\ref{ZNDelta}), are
equivalent when both the nucleon and the $\Delta$ are on-shell.
However, only the choice of Eq. (\ref{ZNDelta}) contains no coupling
to the spin-1/2 component of the $\Delta$.

Amplitudes for the cross-box diagrams can be written down similarly.
The loop integrals with $\Delta$ intermediate state are infrared
safe. We use computer package ``FeynCalc'' \cite{Feyncalc} and
``LoopTools'' \cite{Looptools} to carry out the calculation. The
form factors $F_\Delta$ and $H_\Delta$ are necessary for ultraviolet
regulation of the loop integral and we assume simple dipole form
$\Lambda^4_\Delta/(\Lambda_\Delta^2-q^2)^2$ for both of them. We set
$\Lambda_{\Delta}=1$ GeV for $H_{\Delta}$ and $0.84$ GeV for
$F_{\Delta}$. Variations of these cut-offs do not affect
significantly the results.

The values of the coupling constants $g_{i}'s$ can be determined
from the experimental data of $\gamma N\rightarrow \Delta$ at real
photon point. They are simply linear combinations of the
Jones-Scadron form factors \cite{Jones73} $G_{(M,E,C)}$ at
$Q^2=0$,
\begin{eqnarray}
g_{1}&=&K\left(G_{E}(0)-G_{M}(0)\right),\nonumber \\
g_{2}&=&K
\left(-\frac{M_{\Delta}+3M_{N}}{M_{\Delta}-M_{N}}G_{E}(0)-G_{M}(0)\right)
,\nonumber \\
g_{3}&=&\frac{KM_{N}}{M_{\Delta}}\left(-\frac{(M_{\Delta}+M_{N})G_{C}(0)}{M_{\Delta}-M_{N}}
+\frac{4M_{\Delta}^{2}G_{E}(0)}{(M_{\Delta}-M_{N})^{2}}\right),\nonumber \\
\nonumber
\end{eqnarray}
where $K=\frac{3}{2}\frac{M_{N}}{M_{\Delta}+M_{N}}$. With the use of
the most recent experimental values for the $G's(0)$
\cite{pascal07,Maid07},  we obtain $g_{1}=1.91$, $g_{2}=2.63$, and
$g_3=1.59$. These give $\tilde{g}_{1}=0.48$, $\tilde{g}_{2}=0.66$,
and $\tilde{g}_ {3}=0.40$. They differ from those used in
\cite{kondra05} where $G_{C}(0)$ was approximated to be zero.

Only  coupling constants  $h_{i}'s$ remain to be determined. They
can be inferred from the data of $\nu N\rightarrow \mu \Delta$. Many
experimental papers on neutrino induced $\Delta$ production adopt
the notation of Llewellyn-Smith \cite{smith} where the N-$\Delta$
transition is written as
\begin{eqnarray}
&&\langle \Delta^{++}(p')|J^{W}_\mu|p(p)\rangle
=\overline{u}_{\alpha}(p')
[(\frac{C^{V}_{3}}{M_{N}}\gamma^{\lambda}
+\frac{C^{V}_{4}}{M^{2}_{N}}p'^{\lambda}+ \nonumber \\
&&\frac{C^{V}_{5}}{M^{2}_{N}}p^{\lambda})
(q_{\lambda}g_{\mu}^{\alpha}-q_{\mu}g_{\lambda}^{\alpha})\gamma_{5}+C^{V}_{6}g^{\alpha}_{\mu}\gamma_{5}
+(\frac{C^{A}_{3}}{M_{N}}\gamma^{\lambda}+\nonumber \\
&&\frac{C^{A}_{4}}{M^{2}_{N}} p'^{\lambda})
(q_{\lambda}g_{\mu}^{\alpha} -q_{\mu}g_{\lambda}^{\alpha})
+C_{A}^{5}g^{\alpha}_{\mu}
+\frac{C^{6}_{A}}{M_{N}^{2}}p^{\alpha}q_{\mu}]u(p). \label{smith}
\end{eqnarray}

The form factors in Eq. (\ref{ZNDelta}) can be related to the form
factors defined in Eq. (\ref{smith}) by performing a rotation in
isospace and assuming the nucleon and $\Delta$ are both on-shell.
The resulting relations are
\begin{eqnarray}
&h_{1}=&-\frac{\beta
C^{A}_{4}(0)}{\sqrt{3}}+\frac{2M_{N}}{M_{\Delta}}\,\frac{\beta
C^{A}_{3}(0)}{\sqrt{3}},\nonumber\\
&h_{3}=&-\frac{M_{N}}{M_{\Delta}}\cdot\frac{\beta
C^{A}_{3}(0)}{\sqrt{3}},\,
h_{2}=\frac{M_{N}^{2}}{M_{\Delta}(M_{\Delta}-M_{N})}\,\frac{\beta C^{A}_{6}(0)}{\sqrt{3}},\nonumber \\
&h_{4}=&-\frac{M_{N}^{2}}{M_{\Delta}^{2}}\frac{\beta
C^{A}_{5}(0)}{\sqrt{3}}-\frac{M_{N}(M_{\Delta}-M_{N})}{M_{\Delta}^{2}}\,\frac{\beta
C^{A}_{3}(0)}{\sqrt{3}}.
\end{eqnarray}

According to \cite{Adler, Schreiner} $C^{A}_{3}$=0, hence
$h_{3}$=0. From the data of $\nu N\rightarrow \mu N\pi$
\cite{Kitagaki90} one can extrapolate the experimental result to
$Q^2$=0 and find $C_{4}^{A}(0)=-0.8$ and $C_{5}^{A}(0)=2.4$
\cite{Hemmert}. One then obtains $h_{1}=-0.263$, and
$h_{4}=-0.458$. Parameter $h_{2}$ can not be determined from the
weak pion production. Fortunately its effect is tiny ($\le
10^{-18}$) and we simply set $h_2=0$. The sensitivity of the
results with respect to the variation of $h_i's$ is found to be
very small.

In Fig. \ref{Fig_delta}, we show the TPE and $\gamma ZE$ corrections
to $A_{PV}$ by plotting $\delta=\delta_{N}+\delta_{\Delta}$, defined
by
\begin{equation}
A_{PV}(1\gamma+Z+2\gamma+\gamma
Z)=A_{PV}(1\gamma+Z)(1+\delta_{N}+\delta_{\Delta}),
\end{equation}
{\it vs.} $\epsilon \equiv [1+2(1+\tau)\tan^2\theta_{Lab}/2]^{-1}$,
where $\theta_{Lab}$ is the laboratory scattering angle and
$\tau=Q^2/(4M^2)$, at four different values of $Q^2=0.1,\, 0.5,
\,1.0, $ and $3.0$ GeV$^2$. $A_{PV}(1\gamma+Z)$ denotes the
parity-violating asymmetry arising from the interference between OPE
and $Z$-boson-exchange, i.e., Figs. 1(a) and 1(b), while
$A_{PV}(1\gamma+Z+2\gamma+\gamma Z)$ includes the effects of TPE and
$\gamma ZE$. $\delta_N$ and $\delta_\Delta$ correspond to the
contribution from the box diagrams with only nucleon  or $\Delta$ in
the intermediate states and are represented by dotted and dot-dashed
lines, respectively. The sum $\delta=\delta_N+\delta_\Delta$ is
given by solid curves. In addition, we also present the contribution
arising from interference between $\gamma ZE$ with
$\Delta$-excitation and OPE, $\delta_\Delta(\gamma Z)$ by short
dashed lines. The difference between dot-dashed lines and short
dashed lines would then correspond to effects of interference
between TPE with $\Delta$-excitation and $ZBE$, i.e.,
$\delta_\Delta(2\gamma)=\delta_\Delta -\delta_\Delta(\gamma Z)$.

\begin{figure}[h,b,t]
\centerline{\epsfxsize 3.5 truein\epsfbox{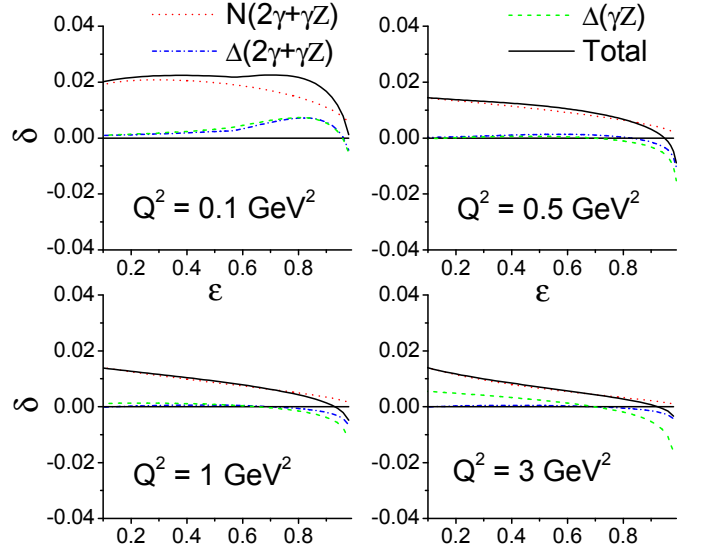}} \vskip-0.7cm
\caption{TPE and $\gamma Z$-exchange corrections with nucleon and
$\Delta$  intermediate states to parity-violating asymmetry as
functions of $\epsilon$ from 0.1 to 0.98 at $Q^2$ = 0.1, \,0.5,\,
1.0, and 3.0\, GeV$^2$.} \label{Fig_delta}
\end{figure}

We see from Fig. \ref{Fig_delta}, that $\delta_\Delta$ (dot-dashed
lines) is very small except at low $Q^2 \sim 0.1$ GeV$^2$ and
$\epsilon \ge 0.7$. Detailed analysis shows that below $Q^2 \sim
0.1$ GeV$^2$,   $\delta_\Delta(\gamma Z)$ starts out as very small
and positive at small $\epsilon$ and increases with $\epsilon$
before turning around at $\epsilon \sim 0.9$  and drops sharply to
become large and negative as $\epsilon$ approaches 1. The peak
around $\epsilon \sim 0.9$ is very sharp and becomes less pronounced
with increasing $Q^2$. In this region of small $Q^2$,
$\delta_\Delta(2\gamma)$ is almost negligible. For $Q^2$ in the
region of $0.1\sim 1.0$ GeV$^2$, both $\delta_\Delta(2\gamma)$ and
$\delta_\Delta(\gamma Z)$ are flat and almost zero until $\epsilon$
increases past 0.8 and results in a small and negative
$\delta_\Delta$ at forward angles.  For $Q^2\ge 3.0$ GeV$^2$,
$\delta_\Delta(\gamma Z)$ starts out around $1\%$ at backward angles
and gradually decreases with increasing $\epsilon$ to cross zero at
$\epsilon \sim 0.7$ before dipping further to large and negative at
extreme forward angles. However, the total effects of the $\Delta$,
$\delta_\Delta$ is small at $Q^2\ge 1.0$ GeV$^2$. These behaviors
differ from those of $\delta_N$ which always stays positive and
decreases with increasing $\epsilon$. At small $\epsilon$,
$\delta_N$ is much greater than $\delta_{\Delta}$ such that the sum
$\delta$ is not much different from $\delta_N$. However, for larger
values of $\epsilon$, $\delta_{\Delta}$  dominates.

\begin{figure}[h,b,t]
\centerline{\epsfxsize 3.5 truein\epsfbox{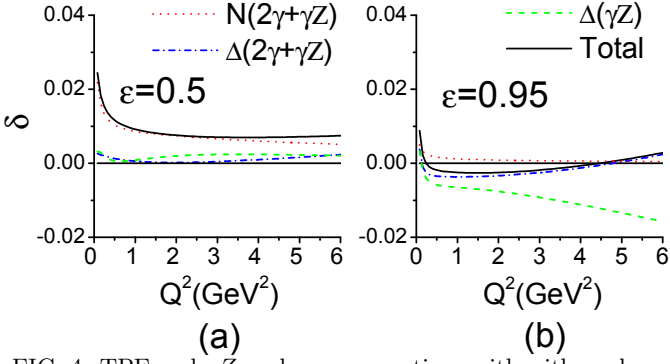}} \vskip-0.5cm
\caption{TPE and $\gamma Z$-exchange corrections with with nucleon
and $\Delta$  intermediate states to parity-violating asymmetry as
functions of $Q^2$ from 0.1 to 6 GeV$^2$ at $\epsilon$ = 0.5 and
0.95.} \label{Q}
\end{figure}

Another way to look at our results is to see the evolution of the
$\delta's$ with respect to $Q^2$ at fixed $\epsilon$ as depicted in
 Fig. {\ref{Q}} for $\epsilon = 0.5$ and $\epsilon = 0.95$.
The notation is the same as in  Fig. \ref{Fig_delta}. We clearly see
that at $\epsilon=0.5$, $\delta_{N}$ is dominant. However, at
$\epsilon\ge 0.95$ where most of the exiting data are taken,
$\delta_{\Delta}$ dominates.

We now proceed to estimate the effects of the TBE  on the values
of strange form factors extracted from HAPPEX, A4 and G0
\cite{expts} experiments. The parity asymmetry is conveniently
\cite{Musolf90} expressed as follows,
\begin{eqnarray}
&&A_{PV}(\rho,\kappa)=A_1+A_2+A_3, \nonumber\\
&&A_{1}=-a\rho\left[(1-4\kappa\sin^{2}\theta_{W})-\frac{\epsilon
G^{\gamma,p}_{E}G^{\gamma,n}_{E}
+\tau G^{\gamma,p}_{M}G^{\gamma,n}_{M}}{\epsilon(G^{\gamma,p}_{E})^2+\tau(G^{\gamma,p}_{M})^2}\right ],\nonumber \\
&&A_{2}= a\rho\frac{\epsilon G^{\gamma,p}_{E}G^{s}_{E}
+\tau G^{\gamma,p}_{M}G^{s}_{M}}{\epsilon(G^{\gamma,p}_{E})^2+\tau(G^{\gamma,p}_{M})^2},\nonumber \\
&&A_{3}=a(1-4\sin^{2}\theta_{W})\frac{\epsilon'
G^{\gamma,p}_{M}G_{A}^{Z}}
{\epsilon(G^{\gamma,p}_{E})^2+\tau(G^{\gamma,p}_{M})^2},
\label{A123}
\end{eqnarray}
where $a= G_{F}Q^2/4\pi\alpha\sqrt{2}$, $\epsilon'=\sqrt{\tau
(1+\tau) (1-\epsilon^2)}$, and $\alpha$ the fine structure constant.
When the parameters $\rho$ and $\kappa$ are set to one, Eq.
(\ref{A123}) reduces to the expression obtained in tree
approximation. The linear combination of the strange form factors
$G^{s}_{E}+\beta G^{s}_{M}$, with $\beta=\tau G^{\gamma,
p}_{M}/\epsilon G^{\gamma,p}_{E}$ has been extracted from $A_2$ in
Eq. (\ref{A123}). Deviations of $\rho$ and $\kappa$ from one
represent all possible higher-order radiative corrections, including
vertex corrections, corrections to the propagators, and $\gamma
Z-$exchange etc.

 The latest PDG values \cite{PDG2008} give $\rho = 0.9876,
\kappa = 1.0026$. To avoid double counting, one should remove the
effect of the $\gamma Z$ box diagram. The values $\delta \rho =
-3.7\times 10^{-3}$ and $\Delta \kappa = -5.3\times 10^{-3}$ for
this effect used in PDG  were those estimated by \cite{Marciano83}
within the zero momentum transfer $Q\equiv 0$ approximation scheme,
when the onset scale is set to be 1 GeV. Consequently, we will set
the experimental parity asymmetry,
\begin{eqnarray}
A_{PV}^{(Exp)}&\equiv &A_{PV}(1\gamma+Z+2\gamma+\gamma Z),
\nonumber\\&=&A_{PV}(\rho',\kappa')(1+\delta_{N}+\delta_{\Delta}).
\label{A-corrected}
\end{eqnarray}
where $\rho'=\rho-\Delta\rho$ and $\kappa'=\kappa-\Delta\kappa$.
We can then determine $A_{PV}(\rho',\kappa')$ and extract the
strange form factors from the resultant $A_2$.

We introduce $\delta_G$
\begin{equation}\overline{G}_E^s+\beta\overline{G}_M^s=(G_E^s+\beta
G_M^s)(1+\delta_G), \end{equation} to quantify the effects of TBE,
where $G_E^s+\beta G_M^s$ and
$\overline{G}_E^s+\beta\overline{G}_M^s$ correspond to the strange
form factors extracted from $A_{PV}(\rho,\kappa)$ and
$A_{PV}(\rho',\kappa')$, respectively.
 From Eq. (\ref{A123}) we then obtain
\begin{equation}
\delta_G=\frac{A^{Exp}_{PV}(\frac{\Delta\rho}{\rho}-\delta)+4a\rho\sin^{2}\theta_{W}
\Delta\kappa-A_{3}\frac{\Delta\rho}{\rho} }{A^{Exp}_{PV}-A_{0}},
\label{deltaG}
\end{equation}
where $A_{0}=A_{1}+A_{3}$.

 We present our results for $\delta_{N}$,
$\delta_{\Delta}$, their sum $\delta$, and $\delta_G$ in Table I for
HAPPEX, A4, and G0 experiments. We also list the corresponding
values, labelled as $\delta_{0}$, in  Table I as would be obtained
in \cite{Marciano83} within $Q\equiv 0$ approximation scheme such
that $\delta_{G}=0$ if $\delta=\delta_{0}$. In other words,
difference between $\delta$ as we obtain and $\delta_0$ represents
the possible $Q^2$-dependence neglected in the estimation of
\cite{Marciano83}.
\begin{table}[htbp]
\begin{tabular}
{|c|c|c|c|c|c|c|}
\hline  & I & II & III & IV & V &VI\\
\hline $Q^2(GeV^2)$& 0.477  & 0.109   & 0.23  & 0.108 &0.232& 0.410\\
\hline $\epsilon$  & 0.974 & 0.994   & 0.83 & 0.83 &0.986&0.974 \\
\hline $\delta_{N}(\%)$  & 0.25& 0.34& 0.86  & 1.30 &0.288&0.275 \\
\hline $\delta_{\Delta}(\%)$ & -0.59& -1.53&0.21 & 0.66 &-0.90&-0.60\\
\hline $\delta (\%)$ & -0.34 &-1.19 & 1.07 &1.96 &-0.61&-0.30\\
\hline $\delta_{0} (\%) $ & 1.03 &2.62 &1.51 &3.13 &1.82&1.417\\
\hline $\delta_G(\%)$ &-25.52 &-75.23& -2.76& -2.27& 13.12&20.62\\
\hline
\end{tabular}
\caption{The corrections $\delta_G$ to $G_E^s+\beta G_M^s$ for
HAPPEX, A4, and G0 experiments. (I, II), (III, IV), and (V, VI)
refer to the HAPPEX,  A4,  and  G0 data, respectively
\cite{expts}.} \label{tab2}
\end{table}

All experimental data included in Table I were obtained in near
forward directions with $\epsilon\ge 0.8$. More specifically, the
HAPPEX and G0 data given in the columns (I, II) and (V, VI) were
taken at extremely forward angles with $\epsilon\approx 0.97$. It is
seen that in this region $\delta_{\Delta}$ completely dominates over
$\delta_{N}$ with opposite sign and pushes the total TBE effects
$\delta=\delta_{N}+\delta_{\Delta}$  away from the zero momentum
transfer approximation values $\delta_{0}$. Thus the large
corrections to $\delta_{G}$ we found in \cite{zhou07} for HAPPEX
data in columns (I, II) with only effect of  nucleon intermediate
states considered, are now enhanced from $(-14.6\%, -45.1\%)$ to
$(-25.5 \%, -75.2 \%)$, respectively. Similarly, for the case of G0
data given in (V, VI), $\delta_G$ increases from $(7.7\%, 13.6 \%)$
to (13.1\%, 20.6\%), respectively. However, for A4 data listed in
columns (III, IV) taken at smaller $\epsilon=0.83$ ,
 $\delta_{\Delta}$ is smaller than $\delta_{N}$ with same sign such that
$\delta_{\Delta}$ brings $\delta$ close to $\delta_{0}$ and results
in smaller $\delta_G$'s.

In summary, we calculate TBE effect with $\Delta$ excitation to the
parity-violating asymmetry of elastic $ep$ scattering in a simple
hadronic model. $\Delta$'s contribution $\delta_\Delta$ is, in
general, much smaller compared with the nucleon contribution
$\delta_N$ in magnitude except at extreme  forward angles.
$\delta_\Delta$ stays very small and flat at backward angles but
turn negative abd large at extreme forward angles, in contrast to
$\delta_N$ which always stays positive but decreases monotonically
with increasing $\epsilon$. Accordingly, interplay between the
contributions of TBE with nucleon and $\Delta$ intermediate states
depends strongly on the kinematics. The $\Delta$ effect is smaller
in magnitude than the nucleon contribution with the same sign for
$\epsilon \le 0.8$, as in the case of A4 experiments. The total TBE
effect $\delta=\delta_N+\delta_\Delta$ is hence enhanced and
approaches  closer to the zero momentum transfer approximation value
$\delta_0$ than $\delta_N$. Thus the corrections to $G^{s}_{E}+\beta
G^{s}_{M}$ in A4 experiments decrease to about $-2.5\%$, when the
effect of $\Delta$ excitation is included. On the other hand, at
extremely forward directions with $\epsilon \sim 0.97$, as in the
cases of HAPPEX and G0 experiments, $\Delta$'s effect becomes
negative and dominates over $\delta_N$. The sum $\delta$ then moves
considerably away from $\delta_0$. The combined TBE correction  to
$G^{s}_{E}+\beta G^{s}_{M}$ found in \cite{zhou07} for the HAPPEX
experiments are thereby enhanced to reach  $-25\% \sim -75\%$, while
in the case of  G0 experiments the corrections are in the range of
$13\%\sim 21\%$.

The fact that we find significant contribution from TBE with
$\Delta$ excitation in the very forward angles, where many of the
current experiments are performed, brings up the question of the
inclusion of higher resonances. Naively, one would expect that
$\Delta(1232)$ would give the largest contribution since it is the
most prominent resonance and, besides, higher resonances are
suppressed because of their larger masses. However, only an explicit
microscopic calculation can answer this question. An extension of
the present calculation to include higher resonances is currently
underway. Recent dispersion relation calculation of the $\gamma ZE$
correction to $Q_{weak}$ \cite{gorchtein09} could be used to clarify
this question in the exact forward scattering. However, our results
indicate that $\delta$  depends sensitively with $Q^2$ at low
momentum transfer so whether dispersion relation method of
\cite{gorchtein09} can be extended  to investigate the TBE
correction to   strange form factors remains to be further explored.
Study of TBE effect with the use of GPD as done in \cite{afanasev05}
and \cite{chen04} for TPE effects, will also be very illuminating in
this regard.

We thanks P.G. Blunden for helpful discussions regarding the
numerical part of this calculation. This work is supported by the
National Science Council of Taiwan under grants nos.
NSC096-2112-M033-003-MY3 (K.N and C.W.K.) and NSC096-2112-M022-021
(S.N.Y.) and by National Natural Science Foundation of China under
grant nos 1974118 and 10805009 (H.Q.Z). H.Q.Z. gladly acknowledges
the support of NCTS/North of Taiwan for his visit and the warm
hospitality extended to him by National Taiwan University. K.N. also
acknowledges the support of NCTS/HsinChu of Taiwan for his visit to
Manitoba.

\end{document}